\documentclass[12pt]{article}
\usepackage{amssymb,graphicx}
\usepackage{amsmath}
\usepackage{epsfig}

\def\d{\partial}
\def\l{\left(}
\def\r{\right)}

\newcommand{\be}{\begin{equation}}
\newcommand{\ee}{\end{equation}}
\newcommand{\bea}{\begin{eqnarray}}
\newcommand{\eea}{\end{eqnarray}}
\newcommand{\bg}{\begin{gather}}
\newcommand{\eg}{\end{gather}}
\newcommand{\bseq}{\begin{subequations}}
\newcommand{\eseq}{\end{subequations}}

\newcommand{\Tr}{{\rm Tr}}

\begin{document}
\begin{center}
  {\Large\bf Phases of massive gravity} \\
\medskip
S.L.~Dubovsky$^{a,b}$\\
\medskip
  $^a${\small
Department of Physics, CERN Theory Division, CH-1211 Geneva 23, Switzerland
  }
\medskip
\\
$^b${\small
Institute for Nuclear Research of
         the Russian Academy of Sciences,\\  60th October Anniversary
  Prospect, 7a, 117312 Moscow, Russia
  }
  \end{center}
\begin{abstract}
We systematically study the most general Lorentz-violating graviton mass
invariant under three-dimensional Eucledian group.  
We find that at general values of mass parameters the
massive graviton has six propagating degrees of freedom, and some of them are
ghosts or lead to rapid classical instabilities.  However, there is a number
of different regions in the mass parameter space where massive gravity is
described by a consistent low-energy effective theory with cutoff
$\sim\sqrt{mM_{Pl}}$. This theory is
 free of rapid instabilities and vDVZ discontinuity. Each
of these regions is characterized by certain fine-tuning relations between
mass parameters, generalizing the Fierz--Pauli condition.  In some cases the
required fine-tunings are consequences of the existence of the subgroups of
the diffeomorphism group that are left unbroken by the graviton mass.  We
found two new cases, when the resulting theories have a property of UV
insensitivity, i.e. remain well behaved after inclusion of arbitrary higher
dimension operators without assuming any fine-tunings among the coefficients
of these operators, besides those enforced by the symmetries.  These theories
can be thought of as generalizations of the ghost condensate model with a
smaller residual symmetry group.  We
briefly discuss what kind of cosmology can one expect in massive gravity and
argue that the allowed values of the graviton mass may be quite large,
affecting growth of primordial perturbations, structure formation and,
perhaps, enhancing the backreaction of inhomogeneities on the expansion rate
of the Universe.
\end{abstract}
\section{Introduction}
During the last few years a significant progress in understanding of
properties of massive gravity in four dimensions has been achieved.  One of
the first steps was the implementation of the St\"uckelberg trick in the
Fierz--Pauli theory~\cite{Arkani-Hamed:2002sp}. As a result, a clear unified
picture of the apparently mysterious properties of the massive gravity has
emerged. Namely, the explicitly covariant St\"uckelberg language 
made it clear that the famous van Dam--Veltman--Zakharov (vDVZ)
discontinuity \cite{vanDam:1970vg,Zakharov} and related non-linear effects
\cite{Vainshtein:1972sx}, as well as the hidden sixth polarization mode of the
massive graviton \cite{Boulware:1973my} are all consequences of the fact that
the kinetic term in the Goldstone sector of the Fierz--Pauli theory is
degenerate in the limit $M_{Pl}\to\infty$. As a result, the theory is not well
defined in the limit when gravity decouples and the cutoff energy
 of the Fierz--Pauli theory in the Minkowski background is
extremely low,
\[
\Lambda_{3}\leq (M_{Pl}^2m)^{1/3}\sim (1000~\mbox{km})^{-1}\;,
\]
where $m$ is a graviton mass, which we take of order the current Hubble scale.
It is worth noting that this may not necessarily imply that the Fierz--Pauli
theory is unacceptable phenomenologicaly. Indeed, recently it was
shown~\cite{Nicolis:2004qq} (see also \cite{Dvali:2004ph}) how similar 
problem~\cite{Luty:2003vm,Rubakov:2003zb}
present in 
the context of co-dimension one 
Dvali--Gabadadze--Porrati (DGP) model~\cite{Dvali:2000hr} may
be cured by taking into account the effects of local curvature.  It is unclear
at the moment whether this mechanism can be implemented in the Fierz--Pauli
case.

The further progress in the four-dimensional theories of massive gravity 
is due to idea
\cite{Arkani-Hamed:2003uy}, that instead of Fierz--Pauli mass term, which
completely breaks diffeomorphism invariance of the theory, one can introduce a
graviton mass, breaking only the time reparametrization invariance,
\[
t\to t+\xi^0(t,x)\;,
\]
while keeping the arbitrary time-dependent spatial diffeomorphisms
\be
\label{gcinv}
x^i\to x^i+\xi^i(t,x^i).  \ee The crucial difference with the Fierz--Pauli
case is that in this model the Goldstone sector (consisting of a single scalar
field) is well behaved in the limit when gravity is decoupled.  At the linear
level the resulting Lorentz non-invariant theory (dubbed ``ghost condensate'')
is characterized by two parameters --- the characteristic length and time
scales. Interestingly, it appeares that both of these parameters can be
substantially shorter than the inverse Hubble scale without contradicting to
observations \cite{Dubovsky:2004qe,Peloso:2004ut}\footnote{It was argued in
Ref.~\cite{Frolov:2004vm} that the limits can be much tighter due to efficient
accretion of the ghost condensate by black holes. It is not clear at the
moment, whether the steady state solutions found in Ref.~\cite{Frolov:2004vm}
are the physically relevant ones.}.  It is worth mentioning, however, that in
many respects ghost condensate is very different from what one would call a
theory of massive graviton. In particular, in this model a massless transverse
traceless mode is present in the spectrum.

Finally, very recently it was argued in Ref.~\cite{Rubakov:2004eb},
 that the generic Lorentz-violating graviton mass term, preserving Euclidean
 symmetry of the three-dimensional space\footnote{A possibility
to introduce mass terms of this type was also mentioned in Ref.~\cite{Damour:2002gp}.},
\be
\label{LVmass}
L_m={M_{Pl}^2\over 2}\l m_0^2h_{00}h_{00}+2m_1^2h_{0i}h_{0i}-m_2^2h_{ij}h_{ij}
+m_3^2h_{ii}h_{jj}-2m_4^2h_{00}h_{ii}\r\;, 
\ee 
where $h_{\mu\nu}$ are
perturbations about Minkowski metric, 
also leads to the theory with well behaved
Goldstone sector in the decoupling limit. More precisely, it was shown to be true
 for $m_0=0$ and provided the other
masses satisfy certain positivity conditions.  As a result, this mass term does
not lead neither to ghosts, nor to low strong coupling scale in the flat
space. Unlike the ghost condensate model, the spectrum of
the theory has a mass gap, and describes five propagating degrees of freedom.
Still, the vDVZ discontinuity is absent in this model and the cutoff scale is
given by \be
\label{cutoff}
\Lambda_2\sim\sqrt{m M_{Pl}}\;\sim (0.1 \mbox{mm})^{-1}\;.
\ee
However, one may be worried by the need to satify the fine-tuning relation
\[
m_0^2=0
\]
 in this theory.  This condition is remininscent of the Fierz--Pauli
fine-tuning and may be a signal of the presence of the analogue of the
Boulware--Deser mode and related instabilities. 

This result indicates that the ghost condensate is unlikely to be a single
consistent infrared modification of gravity and calls for the further
systematic study of the Lorentz-violating graviton mass. This is the main task
of the current paper.

We start in section \ref{goldstones} with the covariant formulation of the
 Lorentz-violating massive gravity.  For this purpose we use a
 St\"uckelberg-like formalism. This formalism is a slightly more flexible
 version of one used in the Lorentz-invariant
 case~\cite{Arkani-Hamed:2002sp}. As in the Lorentz-invariant case, graviton
 mass generically leads to the emergence of four new dynamical degrees of
 freedom --- Golsdtone bosons corresponding to broken reparametrizations of
 four coordinates.  An example of ghost condensate hints that the Goldstone
 sector can be made well behaved in the limit when gravity is decoupled by
 requiring that some residual subgroup of the diffeomorphism group is
 left. However, the group of diffeomorphisms has too many subgroups, so the
 case by case consideration of all possible patterns of symmetry breaking
 seems hopeless.

Instead, in section \ref{explicit} we adopt the bottom-up approach and study
the spectrum of the Lorentz-violating massive gravity at general values of
masses $m_i$ in the decoupling limit $M_{Pl}\to\infty$. 
 In agreement with expectations of Ref.~\cite{Rubakov:2004eb} we
find that at non-zero $m_0^2$ theory generically has either 
ghosts or
exhibits rapid classical instabilities. However, we found  a number of other
fine-tuning relations ensuring the absence of pathologies in the Goldstone sector
at the one-derivative level.
 Actually, the very possibility of the other choices was expected,
as we know that  region with $m_0^2>0$ and all other masses equal to zero
correspond to the ghost condensate model with a well behaved
Goldstone sector.  What is interesting is that one obtaines the well behaved 
Goldstone sector under weaker assumptions as well.

In section \ref{higherderiv} we discuss the effects of higher derivative terms.
We argue, 
that, generically, one should assume an infinite number of fine-tunings
in higher order terms to preserve the above attractive properties of the low-energy
theory. It is worth stressing that this does not imply
 that these fine-tunings cannot be realized in some specific UV completions of
massive gravity. This only means that a {\it generic} UV completion would result in a massive
gravity theory with cutoff scale much lower than $\Lambda_2$.

However, we found two new cases when all the required 
fine-tunings can be ensured by the residual reparametrization symmetry. In both cases
this symmetry is a subgroup of the residual symmetry present in the ghost condensate.
Namely, in one case massive gravity is invariant under
 {\it time independent} spatial diffeomorphisms
\[
x^i\to x^i+\xi^i(x)
\]
and in other under time-dependent shifts of spatial coordinates
\be
\label{intxxt}
x^i\to x^i+\xi^i(t)\;.
\ee
Both theories inherit from the ghost condensate the  property
of UV insensitivity. Namely, they remain 
well behaved under inclusion of all possible higher derivative operators
suppressed by the cutoff $\Lambda_2$ and
compatible with the residual reparametrization symmetry, 
without assuming any fine-tuning relations between their coefficients.

Some of other properties of these massive gravities
 are also similar to  that of the ghost condensate.
For instance, all three models have a scalar degree of freedom with peculiar dispersion relation
\[
\omega^2\propto p^4\;.
\]
However, in many respects two new models are very different from the ghost condensate.
For instance, the tensor graviton mode has non-zero mass in case of residual symmetry
(\ref{intxxt}).
 
In section~\ref{vDVZ} we discuss effects due to mixing with gravity, and prove that
vDVZ discontinuity is absent in the massive gravity models discussed here (both fine-tuned 
and UV insensitive).

We conclude in section~\ref{final} with a brief 
discussion of cosmology in massive gravity. We provide an example of
a theory with massive tensor mode,
where flat homogeneous cosmological solutions are identical to those in the
Einstein theory. This implies that the graviton mass can be substantially larger than the
current Hubble value. Consequently, graviton mass may modify in the interesting way
the dynamics of primordial perturbations and structure formation or, even, gravitational
dynamics at the (super)galactic scales. Even more intriguingly, there
is a possibility, that the presence of small scale inhomogeneities
may significantly affect the expansion rate in massive gravity, potentially providing a
 link between the onset of the cosmological acceleration and epoch of structure formation.
\section{St\"uckelberg trick for Lorentz-violating massive gravity}
\label{goldstones}
In the most general sense by massive gravity we understand any theory described
by the following action
\be
\label{fullaction}
S=-M_{Pl}^2\int d^4x\sqrt{-g}R+\int d^4x\sqrt{-g}F\;,
\ee
where the first term is a usual Einstein--Hilbert term and $F$ is, generally
speaking, an arbitrary function of metric components, their derivatives, and 
coordinates $t$, $x^i$ ($i=1,2,3$) itself. One can add
some matter fields to this system, which
 we assume to be minimally coupled to the metric.
In the current
 paper we will limit our consideration to the case when theory (\ref{fullaction})
admits empty Minkowski space-time as one of the  solutions to the field equations. Also, 
following Ref.~\cite{Rubakov:2004eb}, we assume
that a low-energy effective theory for small perturbations
around this solution is invariant under the Eucledian transformations of the three-dimensional
space coordinates $x^i$. In particular, this implies that function $F$ has no explicit
dependence on the coordinates. Furthemore, we assume that function $F$ depends on a single energy
scale $\Lambda$.

Then, the quadratic theory near Minkowski background is described by the usual
 linearized Einstein--Hilbert action plus mass term of the form
 (\ref{LVmass}), where the overall mass scale $m$ is related to the parameters
 $\Lambda$ and $M_{Pl}$ as
\[
\Lambda=\sqrt{m M_{Pl}}\;.
\]
 
The key issue we address in the current paper is what conditions function $F$ should
satisfy in order dynamics
of metric perturbations near Minkowski background could be
 described by a low-energy effective
field theory with cutoff $\Lambda$.
It was shown
in Ref.~\cite{Arkani-Hamed:2002sp}, that this is not possible in the Lorentz-invariant
case.
The convenient tool allowing  to adress this issue is the
St\"uckelberg formalism.
Our version of this formalism is essentially equivalent to that used in 
Ref.~\cite{Arkani-Hamed:2002sp}. Let us nevertheless briefly describe it, both
to make our presentation more self-contained and because we are using slightly 
more flexible
version of the formalism to allow for Lorentz non-invariant situation.

The first step is to introduce four scalar fields $\phi^\mu$ ($\mu=0,\dots,3$). These are
Goldstone bosons corresponding to four broken diffeomorphisms
\[
x^\mu\to \xi^\mu(x)\;.
\]  
The dynamics of these fields is described by some sigma-model action. 
We assume that this action is characterized by a single energy scale $\Lambda$. The main
requirement to this action is that when coupled to gravity it provides the
following field configuration as a solution to the field equations 
\be
\label{fieldvac}
\phi^\mu=\Lambda^2x^\mu
\ee
\be
\label{metricvac}
g_{\mu\nu}=G_{\mu\nu} \;,
\ee 
where $G_{\mu\nu}$ is some fixed metric (Minkowski metric in our case).  
Provided such a solution exists, one can write
\begin{gather}
g_{\mu\nu}=G_{\mu\nu}+h_{\mu\nu}\\
\phi^\mu=\Lambda^2 x^\mu+\pi^\mu
\end{gather}
and
study dynamics of small scalar and metric perturbations $\pi^\mu$, $h_{\mu\nu}$. This, by
definition, will be a certain version of massive gravity in the space-time
with metric $G_{\mu\nu}$. Indeed, given such a theory, one can fix the unitary
gauge $\pi^\mu=0$ and recover the previous non-covariant formulation
(\ref{fullaction}) of massive gravity\footnote{Note, 
that field equations following from the variations
of the Goldstone fields are consequences of the other equations 
in the unitary gauge.}. 

Inversely, given action
(\ref{fullaction}) one always can make it explicitely covariant reintroducing
the Goldstone fields $\phi^\mu$. Namely, everywhere in the action (\ref{fullaction})
one replaces metric $g_{\mu\nu}$ with its gauge transform
\be
g_{\mu\nu}(x)\to \hat{g}_{\mu\nu}\equiv {\d Y^{\lambda}(x)\over\d x^\mu}
{\d Y^{\rho}(x)\over\d x^\nu}g_{\lambda\rho}\l Y(x)\r\;.
\ee
Here $Y^\mu$ are  component functions of the diffeomorphism
\[
\hat{Y}:x^\mu\to Y^\mu(x)\;.
\]
Analogously, in the presence of matter fields $\psi$ one replaces them everywhere
with their images under gauge transformation.
Clearly, fields $Y^\mu$ appear only in the non-covariant part of the action.
In particular, no direct coupling between these fields and 
matter arises, provided the action for matter is
covariant.

Then action (\ref{fullaction}) considered as a functional of the metric $g_{\mu\nu}$ and 
fields $Y^{\mu}$ is diffeomorphism 
invariant under conventional transformations of the metric,
accompanied by the following transformations of the fields $Y^\mu$,
\be
\label{Ytrans}
Y^\mu\to (\xi^{-1}\circ\hat{Y})^\mu\;,
\ee
where $\xi$ is a change of coordinates and $\circ$ is  a natural multiplication
of two diffeomorphisms.
Transformation law (\ref{Ytrans}) implies that component functions
 $Y^\mu$ do not transform as scalars under changes of coordinates.
However, the transformation law for the component functions of the inverse
mapping $\hat{Y}^{-1}$ is that of the scalar fields,
\[
\l\hat{Y}^{-1}\r^\mu\to \l \hat{Y}^{-1}\circ\xi\r^\mu\;.
\]
These component functions are precisely the Goldstone fields $\phi^\mu$ introduced 
earlier,
\[
\phi^\mu\equiv \l \hat{Y}^{-1}\r^\mu\;.
\]

Now, imagine that massive gravity in the unitary gauge (\ref{fullaction})
has a residual invariance under diffeomorphisms $\xi_H$ from a certain subgroup ${\cal H}$ 
of the whole group of diffeomorphisms. 
In covariant formalism this implies that the theory is invariant under the following
 transformations  of component fields $Y^\mu$,
\be
Y^\mu\to(\hat{Y}\circ\xi_H)^\mu\;.
\ee
In terms of scalar Goldstone fields $\phi^\mu$
this implies that a group ${\cal G}$ of {\it internal global} symmetries of the Goldstone
action contains group ${\cal H}$ as its subgroup, i.e. transformations of the type
\be
\phi^\mu\to \xi^\mu_H(\phi)
\ee
are symmetries of the Goldstone action.
One could  think that it is impossible to construct non-trivial actions with such
symmetries for large enough groups ${\cal H}$. However, the St\"uckelberg
trick described above allows to construct a lot of such actions starting 
from the massive gravity in the unitary gauge.

If, in addition, a group of internal symmetries ${\cal G}$ contains the common subgroup
with the group of isometries  of the
metric $G_{\mu\nu}$, then
 vacuum (\ref{fieldvac}), (\ref{metricvac})
is invariant under the diagonal combination of these two subgroups.

To finish this general discussion the following comment is in order. One may
question the assumption that the massive gravity contains just metric and
Goldstone bosons. Indeed, in the case of gauge fields Higgs mechanism implies
the presence of at least one additional scalar field --- the radial mode of
the Higgs field. Presumably, the successful implementation of the Higgs mechanism in
gravity (i.e., a theory of massive graviton with UV cutoff at the Planck
scale) would also require some extra fields. Unfortunately, such a theory is
not constructed yet\footnote{See, e.g., Ref.~\cite{Gripaios:2004ms} for a recent attempt in
this direction, making use of the vector field.}.  
In this situation, the best we can do is to
introduce the minimal set of fields which is required by the
consistency. Then there is nothing wrong in solving the resulting non-linear
equations, provided  one restricts oneself to the classical solutions
tractable within the low-energy effective theory. Tractable means here, that
the theory of small perturbations around the solution has no pathologies leading to rapid
instabilities and has large enough cutoff
scale. Solutions that do not satisfy these criteria cannot
 be studied without more detailed knowledge of the UV completed theory. 

Let us illustrate the above general discussion by some concrete examples.
For instance, the Fierz--Pauli theory corresponds
to the choice of the Minkowski metric as a background metric $G_{\mu\nu}$,
and the four-dimensional 
Poincare group (i.e., the whole group of isometries of the Minkowski space-time)
as a group\footnote{Strictly speaking, in this way one
obtains some general Lorentz-invariant theory of massive graviton. To ensure
that the graviton mass has the Fierz--Pauli form, the sigma-model action
should satisfy additional fine-tuning conditions.}  $\cal{G}$.

Ghost condensate corresponds to the degenerate situation when action does not depend on the
fields $\phi^i$ at all. Formally, one can say that in this case
a group of internal symmetries $\cal{G}$ is a subgroup of the
 diffeomorphism group shifting time  by a constant and transforming spatial coordinates
in the arbitrary way.  

The Lorentz-violating massive gravity considered in Ref.~\cite{Rubakov:2004eb} 
is also a theory of massive gravity in
the Minkowski space-time.
It corresponds to the choice of the subgroup of the 
4d Poincare group generated by space-time shifts 
and rotations of space coordinates as a group of internal symmetries.
Then the most general
 Goldstone action has the following form at the one-derivative level
\be
\label{partinv}
\int d^4x\Lambda^4\sqrt{-g}F(X,Y^{ij},V^i,Z)
\ee
 where $X$, $Y^{ij}$, $V^i$ and $Z$ are the following scalar quantities
\begin{gather}
X=g^{\mu\nu}\d_\mu\phi^0\d_\nu\phi^0/\Lambda^4\nonumber\\ 
Y^{ij}=g^{\mu\nu}\d_\mu\phi^i\d_\nu\phi^j/\Lambda^4\nonumber\\
V^i=g^{\mu\nu}\d_\mu\phi^0\d_\nu\phi^i/\Lambda^4\nonumber\\
Z=\epsilon^{\mu\nu\lambda\rho}
\d_\mu\phi^0\d_\nu\phi^1\d_\lambda\phi^2\d_\rho\phi^3/(\sqrt{-g}\Lambda^8)
\end{gather}
In the rest of the paper by massive gravity (or Lorentz-violating massive gravity)
we imply this particular model.
Note that $Z^2$ can be written as a function of  $X$, $Y^{ij}$, $V^i$, so that different
functions $F$ in Eq.~(\ref{partinv}) may lead to the same Goldstone action.
The Latin indices $i,j=1,\;2,\;3$ are converted using
the 3d Kronecker symbol $\delta_{ij}$ in Eq.~(\ref{partinv}). 
The energy scale $\Lambda$ is a naive
UV cutoff of this model which is assumed to be much lower than $M_{Pl}$. 
As we explain later, in certain circumstances the
actual UV cutoff may be significantly lower.

Clearly, for constant metrics 
$g_{\mu\nu}$  the field configuration (\ref{fieldvac}) is always a solution to
the field equations following from the action (\ref{partinv}). To satisfy the
second condition (\ref{metricvac}) for a Minkowski space-time
one should find two positive
constants $a,\;b>0$
such that the energy-momentum tensor of the sigma-model (\ref{partinv}) is
zero in the background metric 
\[
\eta_{\mu\nu}=diag(a,-b,-b,-b)\;.
\] 
This condition
translates into  a following pair of equations
\be
\label{minkeqs}
2F_XX={2\over 3}F_{Y^{ij}}Y^{ij}=F-ZF_Z
\ee
for
 the values of function $F$ and its
first derivatives 
\[
F_X\equiv{\d F\over \d X}\;,\;\;F_{Y^{ij}}\equiv{\d F\over \d Y^{ij}}\;,\;
\;F_Z\equiv{\d F\over \d Z}
\]
at the point
\[
X=a^{-1}\;,\;\;Y^{ij}=-b^{-1}\delta^{ij}\;,\;\;V^i=0\;,\;\;
Z=a^{-1/2}b^{-3/2}.
\]
It is straightforward to check that for a generic function $F$ one can always
find $a$ and $b$ such that these conditions are met. Then masses 
$m^2_0,\dots,m^2_4$ are combinations of the first and second
derivatives of the function $F$ at this point. 
The overall mass scale $m$ is related to $\Lambda$ as
\[
\Lambda\sim\sqrt{M_{Pl}m}\;.
\]
\section{Quadratic Goldstone action at the one-derivative level}
\label{explicit}
Let us first study the 
Goldsone sector for general values of masses $m^2_i$ in the absence of gravity.  
To obtain the quadratic Goldstone Lagrangian 
one plugs the ``pure gauge''
metric perturbation
\[
h_{\mu\nu}=\d_\mu\pi_\nu+\d_\nu\pi_\mu\;.
\]
in the action (\ref{LVmass}).
The result is~\cite{Rubakov:2004eb}
\begin{gather}
L={M_{Pl}^2}\l 2m_0^2(\d_0\pi_0)^2+
m_1^2(\d_0\pi_i)^2+m_1^2(\d_i\pi_0)^2+
(4m_4^2-2m_1^2)\pi_0\d_0\d_i \pi_i-m_2^2(\d_i\pi_j)^2-\right.\nonumber\\
\left.
(m_2^2-2m_3^2)(\d_i\pi_i)^2\r\;.
\end{gather}
This Lagrangian can be partially diagonalised by
decomposing fields into components corresponding to different irreducible 
representations of the three-dimensional Eucledian symmetry group. 
The vector sector contains transverse part $\pi^T_i$ of the fields $\pi^i$,
\[
\d^i\pi^T_i=0\;.
\]
The high-energy action in this sector depends only on masses $m_1^2$ and 
$m_2^2$ and has the following form 
\be
\label{vector}
L=M_{Pl}^2\l m_1^2(\d_0\pi^T_i)^2-m_2^2(\d_i\pi^T_j)^2\r\;.
\ee
In general, this action is well behaved for
\be
\label{vectorcond}
m_1^2,\; m_2^2>0\;.
\ee
Indeed, if $m_1^2$ is negative, then $\pi_i^T$ become ghosts, while if 
$m_1^2$ is positive and $m_2^2$ is negative then there are classical 
instabilities at all spatial momenta. Points where at least 
one of these 
masses is zero are special. Thus conditions
\begin{gather}
\label{m1}
m_1^2=0\\
\label{m2}
m_2^2=0
\end{gather}
are candidates on the role of fine-tuning relations. At the moment
we proceed under assumption 
that condition (\ref{vectorcond}) is satisfied and discuss later
what happens if one or both of Eqs.~(\ref{m1}), (\ref{m2}) hold.
It is worth recalling~\cite{Rubakov:2004eb} also, that $m_2^2$ is a mass of the transverse-traceless
tensor graviton mode.

The scalar perturbations are $\pi_0$ and
$\pi_L$, where the latter is defined as
\[
\pi_i={1\over  \sqrt{-\d_i^2}}\d_i\pi_L\;.
\]
 Their Lagrangian
has the following form
\begin{gather}
L=M_{Pl}^2\mbox{\Large (}2m_0^2(\d_0\pi_0)^2+m_1^2(\d_0\pi_L)^2+m_1^2(\d_i\pi_0)^2-
(4m_4^2-2m_1^2)\pi_0\d_0\sqrt{-\d_i^2}\pi_L
-\nonumber\\
2(m_2^2-m_3^2)(\d_i\pi_L)^2\mbox{\Large )}\;.
\label{largescal}
\end{gather}
In the Fourier space the field equations following from this Lagrangian are
\be
\label{matrixeq}
M\l
\begin{array}{c}
\pi_0\\
\pi_L
\end{array}
\r=0\;,
\ee
where $M$ is the following $2\times 2$ matrix,
\be
\label{M}
M=
\l
\begin{array}{cc}
\alpha\nu^2+1 & -i\mu\nu\\
i\mu\nu&\nu^2-\beta
\end{array}
\r\;,
\ee
where
\[
\nu={\omega\over p}\;,
\]
\[
\alpha={2m_0^2\over m_1^2}\;\;,\;\mu={2m_4^2-m_1^2\over
  m_1^2}\;\;,\;\beta=2{m_2^2-m_3^2\over m_1^2}\;,
\]
with $\omega$ and $p$ being energy and absolute value of the 
three-momentum correspondingly.
Generically, there are two propagating degrees of freedom in the scalar sector.
Their dispersion relations are determined by the
 condition that the determinant $d$ of the matrix $M$ is zero,
\be
\label{onshell}
d\equiv\alpha\nu^4+\nu^2(1-\beta\alpha-\mu^2)-\beta=0\;.
\ee
There are some special values of the parameters, where the number of degrees 
of freedom in the scalar sector is reduced. It may happen either
when one of the 
roots of Eq.~(\ref{onshell}) goes to infinity, so that the dispersion relation
for the corresponding mode takes form
\be
\label{poption}
p^2=0\;,
\ee
or when one of the roots goes to zero, so that the corresponding dispersion
relation is
\be
\label{woption}
\omega^2=0\;.
\ee
The first option is realized for 
\be
\label{alpha}
\alpha=0
\ee
so that the order of the equation (\ref{onshell}) reduces by one.
This is the case considered in Ref.~\cite{Rubakov:2004eb}.
There is a 
single propagating mode in the scalar sector in this case.
Actually, there is a possibility to reduce the order of the
on-shell equation even further. This happens when
\be
\label{alpha+}
\alpha=0\;,\;\mu^2=1\;.
\ee
In this case there are no propagating degrees of freedom in the scalar sector.
The second option (\ref{woption}) is realized for
\be
\label{beta}
\beta=0\;.
\ee
Both scalar degrees of freedom have zero energy and
do not propagate if
\be 
\label{beta+}
\beta=0\;,\;\mu^2=1\;.
\ee
Finally, it may happen that  one of
the fine-tuning relations (\ref{alpha}), (\ref{alpha+}) and one of 
(\ref{beta}), (\ref{beta+}) are satisfied simultaniously. For instance, the 
Fierz--Pauli fine-tuning implies that both conditions
(\ref{alpha+}) and  (\ref{beta+}) hold (with $\mu=1$).

Actually,
we are making things somewhat too simple saying
that the above special cases correspond 
to the reduction in the number of degrees
of freedom. This is literally the case at the quadratic level in the 
limit when gravity is decoupled and in the absence of higher derivative terms
in the Goldstone sector. However, one may be worried that inclusion
of any of these effects strongly modifies the whole picture. This really 
happens in some cases, and more detailed discussion of these effects will
be presented in sections~\ref{higherderiv}, \ref{vDVZ}.
The oversimplified picture is enough, however,
for the main purpose of this section --- to identify regions in the mass space,
where obvious pathologies are absent in the low-energy effective theory.
Let us now perform case by case analysis of different possibilities.
\subsection{Generic graviton masses: six propagating degrees of freedom
 and ghosts or classical instabilities}
\label{generic}
To start with, let us consider the most general case when none of the above
fine-tuning relations holds, so that there are two propagating degrees 
of freedom in the scalar sector, and massive graviton has six polarization modes 
in total.
Let us check whether it is possible then to chose parameters $\alpha$,
$\mu$, $\beta$ in such a way that there are no ghosts or classical
instabilities.

Formally the problem is formulated as follows. 
At any given values of the parameters
$\alpha$, $\mu$, $\beta$ and positive $\nu^2$ 
the Hermitian matrix $M$ has two real eigenvalues
$\lambda_{\pm}(\nu^2)$ given by
\be
\label{eigen}
\lambda_{\pm}={t\pm\sqrt{t^2-4d}\over 2}\;,
\ee
where $t$ and $d$ are trace and determinant of the matrix $M$.
 Also there are two values $\nu_{1,2}^2$ where the
on-shell condition (\ref{onshell}) holds, 
i.e. one of the eigenvalues is zero at this point. 

The question is whether it is possible to find 
values of the parameters such that the following two
conditions are met,
\begin{itemize}
\item
Absence of the classical instability,
\be
\label{noinstability}
\nu_{1,2}^2>0
\ee
\item
Absence of ghosts
\be
\label{noghosts}
\left.
{\d\lambda_{\pm}(\nu^2)\over \d\nu^2}\right|_{\nu^2=\nu_{1,2}^2}>0
\ee\;.
\end{itemize}
Let us prove that this is impossible.
The first observation is that to avoid ghosts it is necessary that each of
the eigenvalues $\lambda_\pm(\nu)$ has one zero. Indeed, if one of them has
two zeros and another none than the derivative in one of the zeros is
necessarily negative. Similar situation happens in a theory of a single scalar field with four
derivative kinetic term. 

Now, note that Eq.~(\ref{onshell}) implies that one of the
conditions for the absence of the classical instability is
\be
\label{noclassical}
{\beta\over\alpha}<0
\ee
Let us first discuss the case $\alpha<0$, $\beta>0$.
Here, in the limit of large $\nu^2$ the determinant of the matrix $M$ is
\[
d=\alpha\nu^4+{\cal O}(\nu^2)<0
\]
Consequently, the smaller eigenvalue $\lambda_-$ is negative
 at large $\nu^2$ in this case. This implies, that derivative of $\lambda_-$ in its zero is negative, and
this zero corresponds to the ghost.

In the another case $\alpha>0$, $\beta<0$ the trace of the matrix $M$,
\be
\label{trace}
t=(\alpha+1)\nu^2+1-\beta
\ee
is positive definite.
This implies, that $\lambda_+$ is also 
positive definite. Consequently, $\lambda_-$ has two
zeroes, and one of them corresponds to the ghost. 
\subsection{Phase $m_0^2=0$}
\label{m0phase}
Let us now consider the case when the only fine-tuning relation which 
holds is (\ref{alpha}). This is the case considered in 
Ref.~\cite{Rubakov:2004eb}. 

For $\alpha=0$ one of the solutions $\nu^2$ to the on-shell condition 
(\ref{onshell}) becomes infinite. 
The only remaining solution is
\be
\label{asol}
\nu_0^2={\beta\over 1-\mu^2}
\ee
Let us first consider the case 
\[
\mu^2>1\;.
\]
Then positiveness of $\nu_0^2$ implies that 
\[
\beta<0\;.
\]
As before, in this case the trace (\ref{trace})
 of the matrix $M$ is positive for all positive $\nu^2$,
so $\lambda_+$ is also positive and $\nu_0^2$ is a zero of $\lambda_-$.
On the other hand the determinant $d$ of matrix $M$ is now equal to
\be
\label{0det}
d=\nu^2 (1-\mu^2)-\beta
\ee
and negative at large $\nu^2$. Consequently, $\lambda_-$ is negative
at large $\nu^2$ and has a negative derivative in its zero $\nu_0^2$. So the 
only dynamical degree of freedom in the scalar sector is a ghost here.

Let us now consider the opposite case
\be
\label{mucond}
\mu^2<1\;.
\ee
Then in order to avoid classical instabilities $\beta$ should be positive,
\be
\label{betacond}
\beta>0\;.
\ee
In this case both determinant (\ref{0det}) and trace (\ref{trace})
are positive at large $\nu^2$. Consequently, both eigenvalues $\lambda_{\pm}$
are positive there as well. This implies, that $\nu_0^2$ is a zero of the 
smaller eigenvalue $\lambda_-$, and that derivative of this eigenvalue in its
zero is positive.

Furthemore, if in addition to Eq.~(\ref{alpha}) one also has
  $\mu^2=1$, then the second root of the dispersion relation
(\ref{onshell}) goes to infinity. The determinant of matrix $M$ is equal to
$-\beta$ and non-zero in this case, so  one can use the inverse of this
matrix as a propagator in the perturbation theory (for $\beta\neq 0$).  
The propagating scalar degrees of freedom and obvious
pathologies are absent in the Goldstone sector in this case.  If $\beta$ is
equal to zero, then both fine-tuning relations (\ref{alpha+}) and
(\ref{beta+}) hold and we are in a situation generalizing that in the Fierz--Pauli
theory. Here the quadratic part of the Goldstone sector is zero, so one cannot
organize perturbation theory in this sector in the limit when gravity is
decoupled.

Finally, there is a possibility that fine-tuning conditions (\ref{alpha}) and
(\ref{beta}) are satisfied simultaneously. Here the second mode in the scalar
sector has dispersion relation of the form (\ref{woption}) and is not
propagating as well. There are no obvious pathologies in the Goldstone sector
in this case, provided $\mu^2\neq 1$.

To summarize, Goldstone sector is free of obvious pathologies in the high-energy limit at
$m_0^2=0$,
provided the graviton masses satisfy the following conditions
\begin{gather}
\label{0phase}
m_0^2=0,\;\;m_1^2,\;m_2^2>0,\;\;m_1^2>m_4^2>0,\;\;m_2^2> m_3^2\;\nonumber\\
{\mbox{or}}\nonumber\\
m_0^2=0,\;\;m_1^2,\;m_2^2>0,\;\; (m_1^2-m_4^2)m_4^2=0,\;\;m_2^2\neq m_3^2 \nonumber\\
{\mbox{or}}\nonumber\\
m_0^2=0,\;\;m_1^2,\;m_2^2>0,\;\; m_2^2=m_3^2,\;\;(m_1^2-m_4^2)m_4^2\neq 0\;.
\end{gather}
In the bulk of this phase one of the modes in the scalar sector has dispersion relation
(\ref{poption}) and is not dynamical, so in total
there are three propagating  degrees of freedom in the Goldstone sector.

 We will discuss later
what happens when
either $m_1^2$ or $m_2^2$ is zero. Already now it is clear that if in any of 
the above situations one has
\[
m_2^2=0\;,
\]
then nothing changes in the scalar sector, while both vector degrees of 
freedom become non-dynamical and tensor mode is massless.

The above conditions are in agreement with those found in
Ref.~\cite{Rubakov:2004eb}.  The difference is that we also studied here what
happens on the boundaries of the region found in that work.  Another
difference is that an extra condition $4m_2^2>m_4^2$ was imposed there.  We
will discuss the origin of this extra condition and whether it is really
necessary in section \ref{vDVZ}.
\subsection{Phase $m_2^2=m_3^2$}
\label{m23phase}
In this subsection we proceed under assumption that both conditions
(\ref{vectorcond}) are satisfied, and study what happens if $m_0^2$ is 
non-zero and fine-tuning relation (\ref{beta}) holds. Determinant $d$  of 
the matrix $M$ takes the following form in this case
\be
\label{d23}
d=\nu^2(\alpha\nu^2+1-\mu^2)
\ee
while its trace $t$ is
\be
\label{t23}
t=(\alpha+1)\nu^2+1\;.
\ee
One of the solutions of the on-shell equation (\ref{onshell}) is $\nu^2=0$ 
and does not propagate. Note, that this mode always corresponds to a zero of the
smaller eigenvalue $\lambda_-$, because trace (\ref{t23}) is positive
at $\nu^2=0$. The second solution of the on-shell equation is
\be
\label{nu23}
\nu_{23}^2={\mu^2-1\over\alpha}\;.
\ee
Now, for $\alpha<0$ the determinant (\ref{d23}) is negative at large $\nu^2$, 
and, consequently, $\lambda_-$ is also negative there, while $\lambda_+$ is positive. 
The only possibility to have $\nu_{23}^2$ positive in this case is
if $\nu_{23}^2$ is also a zero of $\lambda_-$, and if the derivative of 
$\lambda_-$ is negative at $\nu_{23}^2$. Hence, one inevitably has classical 
instability or ghost for $m_2^2=m_3^3$ and $m_0^2<0$
(assuming conditions (\ref{vectorcond}) are satisfied). 

For $\alpha>0$ both trace (\ref{t23}) and
determinant (\ref{d23}) are positive at large $\nu^2$, and
consequently $\lambda_+$ and $\lambda_-$ are also both positive there.
Therefore, provided $\nu_{23}^2$ is positive, it is a zero of $\lambda_-$ 
in this case, the sign of the corresponding derivative is positive and
 there are no obvious pathologies in the Goldstone sector.

We have already discussed what happens if $\alpha=0$. Another possibility is 
that $\mu^2=1$, i.e. fine-tuning relation (\ref{beta+}) is realized. 
In this case, there are also no obvious pathologies in the Goldstone sector.
To summarize, there is a phase
\begin{gather}
\label{23phase}
m_2^2=m^2_3>0\;,\;\; m_1^2>0\;,\;\;m_0^2> 0\;,\;\;m_4^2> m_1^2\;
\mbox{or }
m_4^2< 0 \nonumber\\
{\mbox{or}}\nonumber\\
m_2^2=m^2_3>0\;,\;\; m_1^2>0\;,\;\;(m_4^2-m_1^2)m_4^2=0\;,\;\; m_0^2\neq 0\;,
\end{gather}
where Goldstone sector is well defined in the limit when gravity decouples.
In the bulk of this phase one of the modes in the scalar sector has dispersion relation
(\ref{woption}) and is not dynamical, so in total
there are three propagating  degrees of freedom in the Goldstone sector.

As above, nothing changes in the scalar sector if one takes in additon
$m_2^2=0$ so that there are no propagating degrees of freedom in the vector
sector and tensor mode is massless.
\subsection{Phase $m_2^2=0$}
Let us discuss now what happens when mass $m_2^2=0$, so that gradient terms
for the vector modes vanishes and tensor modes are massless. The above
discussion has already covered this case for positive $m_1^2$. If, in addition 
to $m_2^2$, $m_1^2$ is also zero, then the whole quadratic Lagrangian in
the vector sector is zero in the limit when gravity decouples, so 
the Goldstone sector is not well behaved in this limit (the only exception
is the ghost condensate case, when there are unbroken time-dependent 
spatial diffeomorphisms, so that  Goldstones $\pi^i$ do not show up at all).
So the only remaining possibility is that $m_1^2$ is negative,
\be
\label{negm1}
m_1^2<0
\ee
This is possible because vector modes have zero energy and
do not propagate for $m_2^2=0$.

The analysis of the scalar sector proceeds now very similar to what has
been done above.
The only difference
is that the condition that there are no ghosts in the scalar sector is now
opposite to (\ref{noghosts}),
\be
\label{1noghosts}
\left.
{\d\lambda_{\pm}(\nu^2)\over \d\nu^2}\right|_{\nu^2=\nu_{1,2}^2}<0\;.
\ee

We will skip technical details and present just the results. As 
above, there are pathologies in the scalar sector for generic values of
the parameters $\alpha$, $\mu$, $\beta$. 
For $\alpha=0$ the allowed region is 
\begin{gather}
\label{0neg}
m_0^2=m_2^2=0\;,\;\;m_1^2<0\;,\;\;m_3^2< 0\;,\;\;m_4^2< m_1^2\;
\mbox{or }m_4^2> 0 \nonumber\\
\mbox{or}\nonumber\\
m_0^2=m_2^2=0\;,\;\;m_1^2<0\;,\;\;m_3^2=0\;,\;\;m_4^2(m_4^2-m_1^2)\neq 0\nonumber\\
\mbox{or}\nonumber\\
m_0^2=m_2^2=0\;,\;\;m_1^2<0\;,\;\;m_4^2(m_4^2-m_1^2)= 0\,,\;\;m_3^2\neq 0
\end{gather}
For $\beta=0$ one has
\begin{gather}
\label{23neg}
m_2^2=m_3^2=0\;,\;\;m_1^2<0\;,\;\;m_0^2> 0 \;,\;\;m_1^2< m_4^2< 0\nonumber\\
\mbox{or}\nonumber\\
m_2^2=m_3^2=0\;,\;\;m_1^2<0\;,\;\;m_0^2=0\;,\;\;m_4^2(m_4^2-m_1^2)\neq 0\nonumber\\
\mbox{or}\nonumber\\
m_2^2=m_3^2=0\;,\;\;m_1^2<0\;,\;\;m_4^2(m_4^2-m_1^2)= 0\;,\;\;m_0^2\neq 0
\end{gather}
We see, that unlike in the previous cases, fine-tuning
 $m_2^2=0$ alone is never enough to make 
Goldstone sector free of pathologies. So, probably, there is no need to pick out the case 
$m_2^2=0$ into a separate phase.
\subsection{Phase $m_1^2=0$}
\label{m1phase}
The last remaining option is that
\[
m_1^2=0\;,
\]
so that vector modes have dispersion relation of the form
(\ref{poption}) and do not propagate. This case was mentioned in 
Ref.~\cite{Rubakov:2004eb} as a potentially interesting.
In this case the field equation in the
scalar sector still has the form (\ref{matrixeq}) with matrix $M$ being equal 
to
\be
\label{1matrix}
M=
\l
\begin{array}{cc}
2m_0^2\nu^2 & -2im_4^2\nu\\
2im_4^2\nu&2(m_3^2-m_2^2)
\end{array}
\r\;.
\ee
Therefore, the on-shell condition in this case takes the following form,
\be
\label{1oneshell}
d=(4m_0^2(m_3^2-m_2^2)-4m_4^4)\nu^2=0\;.
\ee
If 
\be
\label{1bad}
m_0^2(m_3^2-m_2^2)-m_4^4=0\;,
\ee
then matrix $M$ is degenerate, so that  one cannot use its inverse as a 
propagator in the perturbation theory and Goldstone sector is not well behaved
in the decoupling limit. However, for any other values of parameters Goldstone
sector is well behaved, and there are no propagating vector and scalar modes
in the theory (while tensor mode is massive).
\section{Higher-derivative terms, UV (in)sensitivity and unbroken gauge 
symmetries}
\label{higherderiv}
In this section we continue the study of the Goldstone sector in the limit 
when gravity is decoupled and discuss how the effects of higher derivative 
terms affect the conclusions of  section~\ref{explicit}. Let us, however,
first discuss what happens when  the fine-tuning 
conditions discussed above are slightly violated. Indeed, if these fine-tuning condtions are not
ensured by symmetries, then they are not stable
at the quantum level, so one should check that the low-energy effective theory
remains well behaved if one allows for small violations of fine-tuning relations.
The result of such violation is that the non-propagating degrees of freedom 
become dynamical. Then, there are two different situations depending on whether
the non-propagating degree of freedom had dispersion relation of the form
(\ref{poption}) or (\ref{woption}).

Let us first discuss what happens upon small violation of the dispersion
relation (\ref{woption}). In the vicinity of the point where this dispersion 
relation holds, the corresponding degree of freedom has dispersion relation
of the following form
\be
\label{danger1}
\omega^2=\epsilon p^2\;,
\ee
where $\epsilon$ is a small parameter determined by the distance to the 
fine-tuned point in the mass space. Recall, that the analysis of 
subsection~\ref{generic} implies that in massive gravity
either $\epsilon<0$ leading to classical
instability at all momenta, or $\epsilon>0$ and the corresponding degree of 
freedom is a ghost\footnote{Actually, our analysis does not exclude also a situation, when 
$\epsilon$ is negative and the corresponding degree of freedom is a ghost, i.e. both types of
instabilities are present simultaneously.}. However, in both cases these instabilities do no present
a problem from the viewpoint of the low-energy effective field theory, 
provided $\epsilon$ is small enough. Indeed, for negative $\epsilon$ the 
rate $\Gamma$ for the development of classical instability is smaller than
\[
\Gamma\sim\omega\lesssim |\epsilon|^{1/2}\Lambda
\]
 and can be made arbitrarily small by
chosing a small enough $|\epsilon|$, while keeping the cutoff value
$\Lambda$ constant.  

If $\epsilon$ is positive, then the corresponding degree
of freedom is a ghost, and there is a quantum instability related to the
possibility to produce it from the vacuum together with normal particles of
positive energy.  However, at small $\epsilon$ the maximum (negative) energy
of the ghost quanta available in the regime of validity of the low-energy
effective field theory is very small so that the rate of development of
quantum instability is strongly suppressed by the phase volume.

Definitely,  in both cases there is an instability in the
low-energy effective theory, so rigorously speaking the latter is not well defined.
However, this is not important for any practical purposes provided the time-scale for the
development of instability is long enough. In other words, in all practical situations there
is  an IR cutoff at long time scales (at least at the age of the Universe). 

One could be confused by the fact, that we allow to have ghosts in the 
low-energy effective theory.  Recall, (see, e.g., Ref.~\cite{Cline:2003gs,Holdom:2004yx} 
for a
recent discussion) that ghosts imply quantum instability of the vacuum due to
the possibility of their creation accompanied by normal particles of positive
energy.  In Lorentz-invariant theory the rate of the development of this
instability diverges due to the infinite phase volume, unless one introduces
Lorentz-violating cutoff.  Then the rate is proportional to the value of the
cutoff. In particular, this implies that if the only dimensionfull parameter in
the theory is the cutoff $\Lambda$ itself (as is the case in the Goldstone
sector, where perturbativity requires that the
actual cutoff should be just slightly lower than $\Lambda$), then the rate of
instability is also of order $\Lambda$. One way to make this situation
tractable is to introduce a cutoff at the significantly lower scale. 

However, in
the Lorentz non-invariant situation it may happen that the dispersion relation
is such, that the highest (negative) available energy of the ghost quanta is
very small (see, e.g. dispersion relation (\ref{023mix}) which we discuss
later). It is even possible, that the energy has a maximum at a certain value of momentum 
and goes to zero at higher momenta.
In this case processes with substantial production of energy involve
a large number of ghosts and can be very strongly suppressed even at
relatively large values of the coupling constants. It may be even possible that there is no
need for the explicit Lorentz-violating cutoff
to make the rate of instability finite  --- Lorentz-violating dynamics of the system
itself may play a role of such a cutoff.
 A somewhat degenerate
example of a situation of this type happens in quantum cosmology, where scale factor of
the Universe is a ghost with zero energy.

We believe that these issues deserve further studies. In any case, it is clear that if the
maximum frequency of the ghost quanta is smaller than the inverse 
age of the Universe, then time uncertainty principle makes it nonsensible 
to speak about production of ghosts.

Let us now discuss what happens
when  the dispersion relation of the form (\ref{poption}) is slightly violated.
In this case one has a mode with dispersion relation
\be
\label{danger}
\epsilon\omega^2=p^2\;,
\ee
which is, in our case,
again either indicates the presence of classical instability at
$\epsilon<0$ or a ghost at $\epsilon>0$. Here the low values of momenta,
\[
p\lesssim|\epsilon|^{1/2}\Lambda
\]
are dangerous. Indeed, at higher momenta the presence or absence of instability
depends on the structure of the theory
above the cutoff energy, i.e. out of the domain of applicability of the 
low-energy effective field theory, so one can consistently assume that 
instability is absent.
To eliminate the dangerously rapid
instability at the small values of the momentum one needs
 some IR regulator. This regulator can be provided, for instance, by the finite
size of the system, or by the Hubble expansion (in some similarity to how
Jeans instability is eliminated at large length scales). 
Another option (which, as we discuss in section~\ref{vDVZ}, is realized,
e.g., for $m_0^2=0$ phase) is that mixing with gravity leads to
the following modification of the dispersion
relation (\ref{danger}),
\be
\label{safe}
\epsilon\omega^2=p^2+m^2\;,
\ee
where $m^2$ should be positive.

To summarize, there is an interesting complementarity of
 two situations (\ref{danger1}) and (\ref{danger}). 
Namely, in the first case IR cutoff in
time and UV cutoff in space  are needed to stabilize the theory, while in
the second case one needs IR cutoff in space and UV cutoff in time\footnote{We
are thankful to Riccardo Rattazzi for numerous fruitful discussions of
these stability issues and for the collaboration on the subject.}.

Having discussed what happens upon small departure from the dispersion 
relations (\ref{poption}) and (\ref{woption}) at the one-derivative level, let 
us turn to the effects of the higher derivative terms. 
Generically, they will also make the would be non-dynamical degrees of freedom
 dynamical. For instance, generic higher derivative terms would lead to the
following modification of the 
dispersion relation (\ref{woption})
\be
\label{alaghostcond}
\omega^2=a{p^4\over\Lambda^2}+\dots\;,
\ee
where $a$ is a coefficient of order one.
This is precisely the situation in the ghost condensate model, where such 
modification does not lead to any problems (provided $a$ is positive), because 
the resulting degree of freedom has a healthy (positive) kinetic term.
However, in many of other cases discussed in section~\ref{explicit}
this is not the case and the resulting degree of freedom is a ghost.
Effect of higher derivative terms on the dispersion relation
(\ref{poption}) is even worse. Indeed, one obtains dispersion relation of the
form
\be 
p^2=b{\omega^4\over\Lambda^2}+\dots
\ee
inevitably leading to both quantum and classical instability.

The above discussion implies that the {\it generic} effect of higher
derivative terms is to make the actual cutoff of the theory much lower than
the naive value $\Lambda$. It is worth stressing, that this 
does not imply that one cannot assume that the structure of the theory in the
 UV is such, that coeffcients $a$, $b$ and their higher order analogues
are very small, so that they do not lead to any pathologies in the low-energy
effective theory below cutoff $\Lambda$, due to the same reasons as small departures from
fine-tuning relations at the quadratic level. Clearly, from the
viewpoint of low-energy effective field theory this is a very fine-tuned 
situation, but there is nothing principally wrong in considering such a 
situation. 

While, a priori, there is nothing  wrong  in considering
a fine-tuned situation, it is worth looking for some symmetry
principle
ensuring the desired fine-tunings and thus making massive gravity
well behaved without assuming any fine-tuning relations in the Goldstone
action (besides those, 
following from the symmetries). An example of massive gravity  having this
attractive property of the UV insensitivity is a ghost condensate. This is 
achieved by requiring that there is a residual invariance under time-dependent
spatial reparametrizations (\ref{gcinv}).

 Besides the purely esthetic attractiveness this property of UV
insensitivity may be very important 
from the practical viewpoint as well. Indeed, eventually, we are interested
in the study of the non-trivial curved backgrounds in massive gravity, and 
in the absence of a symmetry principle there is no guarantee, that, say, the
cosmological expansion would not drive a sysetm out of the safe region where
fine-tuning relations hold. 

To illustrate this, let us consider the following
simple choice of function $F$ in the Goldstone action (\ref{partinv}),
\be
\label{simple}
F=P(X)+(\delta_{ij}Y^{ij})^2
\ee
This is just a ghost condensate action coupled to three free scalar fields
$\phi^i$. It is straightforward to check that fine-tuning condition
$m_0^2=0$ implies that 
\be
\label{m0simple}
P_X+2XP_{XX}=0\;,
\ee
and that for a large class of functions $P$ one can get a massive gravity in phase $m_0^2=0$ using
this type of actions for Goldstones.
On the other hand, equation for field $\phi^0$ in the expanding
Universe in this model is the same as in
the case of ghost condensate,
 \be
\d_t\l a^3P_X\d_t\phi^0\r=0\;,
\ee
implying that at late stages of the expansion  a system is driven either
to a point 
\be
\label{triv}
\d_t\phi^0=0\;,
\ee
or to the ghost condensate point
\be
\label{gc}
P_X=0\;.
\ee
(in both cases $Y$ goes to zero). However, it is straightforward
to check that it is not possible to move from the vicinity of  
the point where condition (\ref{m0simple}) is satisfied to points (\ref{triv}),
(\ref{gc}) without crossing  regions where instabilities are not suppressed 
by any small parameter.
 
So, let us try to see whether it is possible to protect fine-tuning relations 
obtained in section~\ref{explicit} by residual reparametrization symmetries.
In each of the phases found there one of the conditions
$m_0^2=0$, $m_1^2=0$ or $m_2^2=m_3^2$ holds. 
Let us see what symmetries can protect 
each of these relations.

Let us start from the relation $m_1^2=0$. It can be protected by one of the 
following residual reparametrization symmetries
\begin{eqnarray}
\label{xtx}
x^i\to x^i+\xi^i(t,x)\\
\label{xt}
x^i\to x^i+\xi^i(t)\\
\label{ttx}
t\to t+\xi^0(t,x)\\
\label{tx}
t\to t+\xi^0(x)
\end{eqnarray}
First symmetry (\ref{xtx}) is just a residual symmetry of the ghost condensate,
so we will not discuss it here. 

At the  linear level the second symmetry (\ref{xt})
implies invariance under arbitrary time-dependent
shifts of the $\pi^i$ fields. In other words, field configuration
\[
\pi^0=0\;,\;\;\pi^i=\pi^i(t)
\]
is a solution of linearized field equations, provided there is a residual reparametrization
symmetry (\ref{xt}). In particular, this implies, that for arbitrary
choice of higher derivative terms, compatible with symmetry (\ref{xt}),
the dispersion relation in the vector sector is (at non-zero $m_2^2$)
\be
\label{xtvector}
p^2\l 1+a{\omega^2\over\Lambda^2}+b{p^2\over\Lambda^2}+\dots\r=0\;,
\ee
where $a$ and $b$ are again some coefficients of order one.
This is in agreement with what we found at the one derivative level, and 
implies that there are no propagating light vector degrees of freedom in this case.

Similarly, in the scalar sector symmetry (\ref{xt}) implies,
that for generic higher derivative terms the on-shell
condition in the scalar sector has the following  form
\be
\label{xtscalar}
 p^2(\omega^2+a{p^4\over\Lambda^2}+\dots)=0\;,
\ee
where the lowest order term is determined
by our explicit analysis in subsection~\ref{m1phase}.
Therefore, similarly to the case of ghost condensate, there is one propagating 
degree of freedom in the scalar sector with peculiar dispersion relation
\be
\label{gcdisp}
\omega^2\propto p^4
\ee
in this case. In order this degree of freedom had a healthy sign before kinetic 
term the derivative of the eigenvalue of matrix (\ref{1matrix}) having zero
at $\nu^2=0$ should be positive at this point. It is straightforward to 
check that this is the case, provided
\be
\label{1deriv}
m_0^2-{m_4^4\over (m_3^2-m_2^2)}>0\;.
\ee
The point $m_3^2=m_2^2$ is rather special --- here both eigenvalues of matrix
(\ref{1matrix}) have square root branchings at $\nu^2=0$. Perhaps,
this case deserves further study.

Thus, insisting on the residual symmetry (\ref{xt}), one obtains a 
generalization of the ghost condensate theory sharing all attractive features 
of the latter. On the other hand this model has a number of new features.
For instance, tensor gravitational waves are massive here.
 Clearly, this theory deserves a separate detailed study.

Symmetry (\ref{ttx}) implies that all masses but $m_2^2$ and $m_3^2$ vanish.
As we discussed in subsection~\ref{m1phase} in this case matrix $M$ is 
degenerate, so that Goldstone sector is not well behaved in the decoupling 
limit. 

Finally, 
symmetry (\ref{tx}) does not allow to put any restrictions on
higher-derivative corrections  to the dispersion relation in the vector 
sector, so it does not allow to avoid fine-tuning. Similarly, in the scalar 
sector it implies that dispersion relation has a form
\[
\omega^2(p^2+a{\omega^4\over\Lambda^2}+\dots)=0\;,
\]
that does not protect from the instabilities.

\begin{figure}[t]
\begin{center}
\epsfig{file=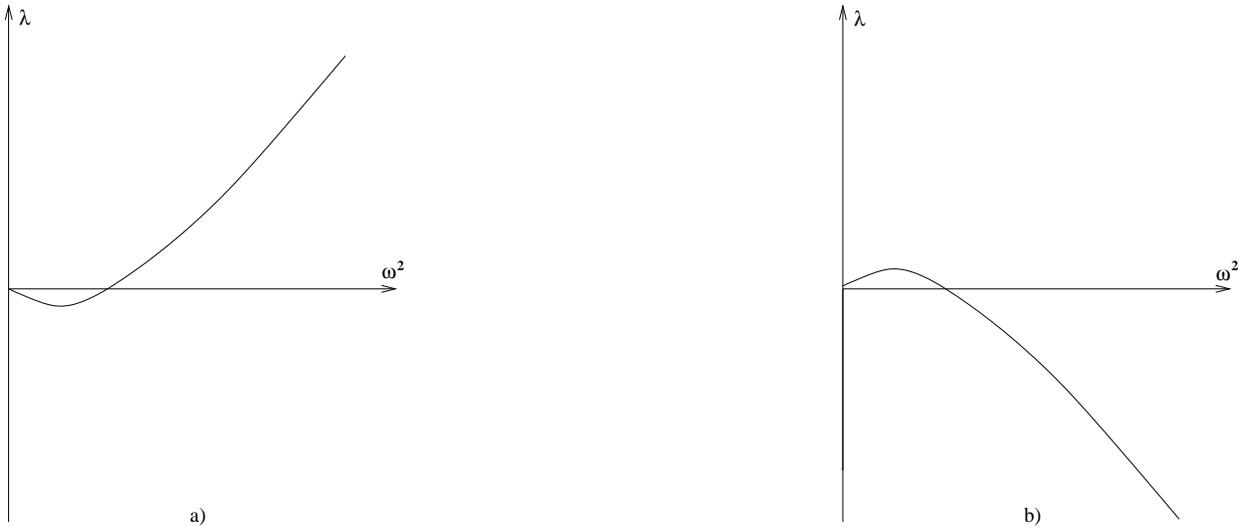,height=7cm}
\end{center}
\caption{Splitting of zeroes of the eigenvalue of the matrix $M$ due to higher-derivative terms
in the presence of residual symmetry (\ref{xx})
 in two different cases: a) $m_1^2\;,\;m_0^2>0$, b) $m_1^2\;,\;m_0^2<0$}
\label{split}
\end{figure}

Let us now discuss symmetries protecting fine-tuning relation $m_2^2=m_3^2$.
Besides symmetry (\ref{xtx}), leading to ghost condensate, this fine-tuning
relation can be protected by a less restrictive residual symmetry
\be
\label{xx}
x^i\to x^i+\xi^i(x)\;.
\ee
Actually, this symmetry implies a stronger condition
\be
\label{234strong}
m_2^2=m_3^2=m_4^2=0\;.
\ee
In the vector sector it protects from any instabilities
in the low-energy effective theory, leading to the dispersion relation
\be
\label{xxvector}
\omega^2(1+a{\omega^2\over\Lambda^4}+\dots)\;,
\ee
provided $m_1^2\neq 0$.
In the scalar sector symmetry (\ref{xx})
implies the following form of the dispersion relation (at the lowest order we are
 using the result of subsection~\ref{m23phase})
\be
\label{xxscalar}
\omega^2(\omega^2+a{p^4\over\Lambda^2}+\dots)=0\;.
\ee
Consequently, we again obtain one propagating degree of freedom in the scalar
sector with dispersion relation (\ref{gcdisp}).
To understand when this mode has a healthy kinetic term let us look at the
 behavior of the eigenvalues of matrix (\ref{M}) when masses $m_2^2$, $m_3^2$ 
and $m_4^2$ are zero. At this point one of the eigenvalues has a double zero
at $\omega^2=0$ while another has no zeros at all. Higher order terms in the
dispersion relation (\ref{xxscalar}) lead to the splitting of the double zero
into two simple zeroes as shown in Fig.~\ref{split} (for negative value of the 
coefficient $a$ in Eq.~(\ref{xxscalar})). Now, for positive $m_1^2$ the 
propagating mode has a healthy kinetic term in the situation shown in the left
panel of Fig.~\ref{split}, while for negative $m_1^2$ in the situation shown in
the right panel of Fig.~\ref{split}. 
Straightforward calculation shows that 
propagating mode is not a ghost provided
\be
\label{xxnoghost}
m_0^2m_1^2>0\;.  
\ee 
This UV insensitive theory resembles ghost condensate
even more, because tensor mode is massless in both cases. Also, in both cases
graviton mass has a form of a gauge fixing term and thus higher order terms (non-linear,
or higher derivative) are needed to see modification of gravity.

Finally, let us discuss whether it is possible to make use of
a residual gauge symmetry to avoid fine-tunings in the phase with $m_0^2$=0.
Besides the symmetry (\ref{ttx}) leading to the degeneracy in the Goldstone 
sector one may try the following symmetry
\be
\label{tt}
t\to t+\xi^0(t)
\ee 
This symmetry implies that 
\[
m_0^2=m_4^2=0\;.
\]
Unfortunately, in the scalar sector this symmetry implies
the dispersion relation  of the form
\be
\label{ttscalar}
p^2(p^2+a{\omega^4\over\Lambda^2}+\dots)=0
\ee
and thus is not enough to obtain a UV insensitive theory.

To summarize, we found two new examples of the UV insensitive massive gravities.
This is the whole phase $m_1^2=0$ and a part of the boundary of $m_2^2=m_3^2$ phase, where
condition (\ref{234strong}) holds.
In both cases the residual symmetry group is a subgroup of the residual 
symmetry of the ghost condensate. Both theories deserve further detailed study.

Clearly, our analysis does not exhaust all possible subgroups of the 
diffeomorphism group. It is worth studying whether other interesting 
possibilities exist.
\section{Mixing with gravity and absence of the vDVZ discontinuity}
\label{vDVZ}
Let us discuss now,
how mixing with gravity may affect the above  conclusions
and show  that the vDVZ discontinuity is absent
if graviton masses are in the ranges found in section~\ref{explicit}.

Let us start from the discussion of the vDVZ discontinuity.
The complete linearized field equations in massive gravity can be presented
schematically as follows
\begin{equation}
\l
{\cal M}_0+{\cal M}_m
\r
\l
\begin{array}{c}
h_{\mu\nu}\\
\pi_\mu
\end{array}
\r
=
\l
\begin{array}{c}
T_{\mu\nu}\\0
\end{array}
\r\;,
\end{equation}
where $T_{\mu\nu}$ is the energy-momentum tensor of matter, we assumed that
there are no sources for the Goldstone fields $\pi_\mu$;  matrix ${\cal M}_0$
is diagonal
\be
\label{M0}
{\cal M}_0=
\l\begin{array}{cc}
M_{Pl}^2{\cal D}_E^2&0\\
0 &\Lambda^4{\cal D}_G^2
\end{array}
\r\;,
\ee
and matrix ${\cal M}_m$ accounts for graviton masses and kinetic mixings 
between Goldstones and gravitons,
\be
\label{Mm}
{\cal M}_m=
\l\begin{array}{cc}
\Lambda^4 {\cal O}& \Lambda^4{\cal D}_m\\
 \Lambda^4{\cal D}_m &0
\end{array}
\r\;.
\ee
Here we assumed that gauge symmetry is fixed in the graviton sector (for 
instance, using harmonic gauge), so that ${\cal D}_E^2$ is the ordinary 
two-derivative graviton kinetic operator following from the Einstein--Hilbert 
action with gauge fixing terms, 
${\cal D}_G^2$ is the two derivative kinetic operator in the Goldstone
sector, ${\cal O}$ is a dimensionless matrix, determining the tensor
structure of the graviton mass matrix, and ${\cal D}_m$ is a one-derivative
operator describing mixing between Goldstones and graviton. For graviton 
mass parameters found in section~\ref{explicit} both operators ${\cal D}_E^2$ and
${\cal D}_G^2$ are invertible at general values of energy $\omega$ and momentum
$p$. Therefore, one can present matrix ${\cal M}$ in the
following form
\be
\label{novDVZ}
{\cal M}={\cal M}_0^{1/2}\l {\mathbf{1}}+
\l
\begin{array}{cc}
{\Lambda^4\over M_{Pl}^2}{\cal D}_E^{-1}{\cal O}{\cal D}_E^{-1}&
{\Lambda^2\over M_{Pl}}{\cal D}_E^{-1}{\cal D}_m{\cal D}_G^{-1}\\
 {\Lambda^2\over M_{Pl}}{\cal D}_G^{-1}{\cal D}_m{\cal D}_E^{-1}&
0
\end{array}
\r
\r
{\cal M}_0^{1/2}\;.
\ee
This presentation makes it manifest, that, far from the 
 on-shell values of energy and momentum found in section~\ref{explicit},
the metric induced by the source $T_{\mu\nu}$ differs from that in the pure
Einstein theory by small corrections, vanishing when
the scale of graviton masses
\[
m={\Lambda^2\over M_{Pl}}
\]
goes to zero. 
In particular, this is true in the Newtonian regime, provided energy
$\omega$ is much larger than the graviton mass scale
\[
m\ll\omega\ll p<\Lambda\;.
\]
Hence, on time and length scales shorter than the inverse graviton mass
one reproduces the usual Newtonian limit of gravity and vDVZ 
discontinuity is absent.

Let us check now whether mixing with gravity can introduce dangerous instabilities
for the ``safe'' phases found in section~\ref{explicit}. The most straightforward way to do 
this is, following Ref.~\cite{Rubakov:2004eb}, to study the spectrum of massive gravity in 
the unitary gauge in each of the cases. Namely, one writes 
\begin{gather}
h_{00}=\psi\\
h_{0i}=u_i+\d_i v\\
h_{ij}=\chi_{ij}+(\d_is_j+\d_js_i)+\d_i\d_j\sigma+\delta_{ij}\tau
\end{gather}
where $\xi_{ij}$ is a transverse-traceless tensor, $u_i$ and $v_i$ are transverse vectors
and other fields are three-dimensional scalars.
Let us start from the vector sector. Here one has the following action~\cite{Rubakov:2004eb}
\be
\label{mixedvect}
L_v=M_{Pl}^2\l s_i\d_0^2\d_j^2s_i-u_i\d_j^2u_i+2u_i\d_j^2\d_0s_i+m_1^2u_iu_i+m_2^2s_i\d_j^2s_i \r\;.
\ee
Dispersion relation following from this action is
\[
m_1^2p^2\omega^2-m_2^2p^2(p^2+m_1^2)=0\;.
\]
Consequently, the only effect of mixing with gravity in the vector sector is
the emergence of the mass gap $m_2^2$.

Let us now consider the scalar sector.
Here, at general values of masses, one has the following quadratic action in the unitary gauge
\begin{gather}
\label{generscal}
L={M_{Pl}^2\over 2}\l 4\l\psi-2\d_0v+\d_0^2\sigma\r\d_i^2\tau+6\tau\d_0^2\tau-2\tau\d_i^2\tau
+m_0^2\psi^2+2m_1^2(\d_iv)^2-\right.\nonumber\\
\left.
m_2^2\l\l\d_i^2\sigma\r^2+2\tau\d_i^2\sigma+3\tau^2\r
+m_3^2\l\d_i^2\sigma+3\tau\r^2-
2m_4^2\psi\l\d_i^2\sigma+3\tau\r
\r\;.
\end{gather}
Let us start our analysis without assuming any fine-tuning relations between masses.
Fields $\psi$ and $v$ enter action (\ref{generscal}) without time derivatives \cite{Rubakov:2004eb}
(the term propartional to
$\d_0v\d_i^2\tau$ in Eq.~(\ref{generscal}) may be written as $v\d_0\d_i^2\tau$).
Field equations resulting from variation with respect to these fields can be written as
\begin{gather}
\label{veq}
v={2\over m_1^2}\d_0\tau\\
\label{psieq}
\sigma={2\over m_4^2}\tau-{3\over\d_i^2}\tau+{m_0^2\over m_4^2}{1\over \d_i^2}\psi\;.
\end{gather}
Plugging expressions (\ref{veq}) and (\ref{psieq}) back into action we
arrive at the following Lagrangian for the two remaining fields $\tau$ and
$\psi$ (we will discuss later what happens if $m_1^2=0$ or $m_4^2=0$),
\begin{gather}
L={M_{Pl}^2\over 2}\l\l{8\over m_4^2}-{8\over m_1^2}\r\d_i^2\tau\d_0^2\tau-4{m_2^2-
m_3^2\over m_4^4}\l\d_i^2\tau\r^2-
6\tau\d_0^2\tau+\l 8{m_2^2\over m_4^2}-2\r\tau\d_i^2\tau-6m_2^2\tau^2+\right.\nonumber\\
\left.
4{m_0^2\over m_4^2}\tau\d_0^2\psi-4{m_0^2\over m_4^4}\l m_2^2-m_3^2\r\tau\d_i^2\psi
-\l m_0^2+{m_0^4\over m_4^4}(m_2^2-m_3^2)\r\psi^2+4m_0^2{m_2^2\over m_4^2}\tau\psi\r\;.
\label{2scalars}
\end{gather}
Let us first consider the phase $m_0^2=0$, when all terms in the second line of Eq.~(\ref{2scalars}) 
vanish. But before actually setting them to zero, the following comment is in order. Straightforward 
inspection of action (\ref{2scalars}) shows, 
that the dispersion relation of the mode which is non-dynamical at 
$m_0^2=0$ takes the following form at small (much smaller than all other masses) non-vanishing $m_0^2$,
\be
\label{smallm0}
m_0^2\omega^2=-{3\over 2}{m_4^4\over\mu_0^2}(p^2+\mu_0^2)\;,
\ee
where parameter $\mu_0^2$ is defined as follows
\[
{4\over m_4^2}-{4\over m_1^2}={3\over\mu_0^2}\;.
\] 
We see, that mixing with gravity automatically provides an IR regulator of the type (\ref{safe})
needed to stabilize the Goldstone sector upon small departure from the surface $m_0^2=0$.

Stability of the action (\ref{2scalars}) in the bulk of the 
$m_0^2=0$ phase was studied in Ref.~\cite{Rubakov:2004eb}. 
As we mentioned in subsection~\ref{m0phase} the only difference of the results of this analysis with 
ours is that extra stability condition
\be
\label{extra}
4m_2^2>m_4^2\;
\ee
was imposed in that work.
If this condition is violated, the term proportinal to $\tau\d_i^2\tau$ enters action (\ref{2scalars})
with negative sign. Depending on other parameters, this may lead to the Jeans-like instability
in the finite region of small momenta and frequencies, $\omega,p\sim m$. Presence of such instability
does not invalidate the use of the low-energy effective theory with cutoff $\Lambda$ for time scales shorter than
$\sim m^{-1}$ and may even lead to
interesting phenomenological consequences. So we do not think that one should a priori disregard mass
values violating condition (\ref{extra}).

Let us see now what are the effects of mixing with gravity at the boundaries
of the phase $m_0^2=0$.  If $m_1^2=m_4^2$, then action (\ref{2scalars})
describes a propagating degree of freedom with the following dispersion
relation (cf. Ref.~\cite{Rubakov:2004eb}) 
\be
\omega^2=(p^2+z\mu_1^2){p^2\over\mu_1^2}+m_2^2\;, \ee where \be
2{m_2^2-m_3^2\over m_4^4}={3\over\mu_1^2}\;,\;\; 4{m_2^2\over m_4^2}-1=3z\;.
\ee
We see, that in this case mixing with gravity invalidates our conclusion
that there are no propagating degrees of freedom in the scalar sector. The new
degree of freedom has a healthy kinetic term and does not lead to the vDVZ
discontinuity provided $m_2^2\neq m_3^2$.  In order to avoid rapid classical
instabilities, one needs, however,
\[
\mu_1^2>0\;,
\]
or, equivalently,
\be
\label{23mcond}
m_2^2>m_3^2\;.
\ee
This condition was impossible to obtain from inspecting 
the Goldstone Lagrangian in the decoupling limit at
$m_1^2=m_4^2$.

If $m_2^2=m_3^2$ the dispersion relation obtained in Ref.~\cite{Rubakov:2004eb} reduces to
\be
\label{023mix} 
\omega^2=\mu_0^2{zp^2+m_2^2\over p^2+\mu_0^2}\;.
\ee
Again we see that mixing with gravity results in the emergence of a new propagating degree of freedom.
This degree of freedom does not lead to any rapid instabilities, provided
\[
\mu_0^2>0\;,
\]
or, equivalently,
\be
\label{04mcond}
m_1^2>m_4^2>0\;.
\ee
As before, this condition is impossible to deduce from the Goldstone Lagrangian alone.
Finally, if in addition to $m_0^2=0$ one has $m_4^2=0$, then Eq.~(\ref{psieq}) implies that
$\tau=0$. Then action in the scalar sector takes the following form 
\be
L={M_{Pl}^2\over 2}(m_3^2-m_2^2)(\d_i^2\sigma)^2\;
\ee
and does not describe any propagating degrees of freedom.
Consequently, mixing with gravity does not bring nothing new in this case.

Let us now discuss effects of mixing with gravity for phase $m_1^2=0$,
protected by the residual gauge symmetry (\ref{xtx}). Here Eq.~(\ref{veq})
implies, that either $\omega=0$, or $\tau=0$. In the latter case, one uses
Eq.~(\ref{psieq}) to eliminate $\psi$ and arrives at the following action for
$\sigma$, 
\be L={M_{Pl}^2\over 2}\l m_3^2-m_2^2-{m_4^4\over
m_0^2}\r\l\d_i^2\sigma\r^2 \;,
\ee 
which does not describe any propagating
fields. So in this case mixing with gravity does not introduce any extra
degrees of freedom.

We believe, that the above discussion is representative enough to demonstrate
what are the possible effects of mixing with gravity. So we leave the detailed
study of the consequences of this mixing in the $m_2^2=m_3^2$ phase and for $m_2^2=0$
with negative $m_1^2$ beyond the scope of the current
paper. Instead, we just consider the point corresponding to the second UV
insensitive theory found in section~\ref{higherderiv}.

Before doing that, let us mention one more possible effect of mixing, that was not
 realized in the phase $m_0^2=0$. Namely, it follows from Eq.~(\ref{2scalars}) 
that in the phase with $m_2^2=m_3^2$ (and non-zero $m_0^2$) a generic
consequence of mixing with gravity is that
 sixth polarization mode of the graviton becomes dynamical. Discussion
in subsection~\ref{m23phase} suggests that this mode may be a ghost or lead to classical
instabilities. Presumably, there
are regions of mass parameters where this is not dangerous.
Our discussion in section~\ref{higherderiv} implies, that what one needs, is that
all quanta of this mode are very soft (cf. dispersion relation (\ref{023mix}))
and, consequently, the corresponding instability rate is very slow\footnote{Note, that
 this may imply additional limitations on the possible values of masses, that are not seen in the
decoupling limit.}.

For point $m_2^2=m_3^2=m_4^2=0$, protected by the symmetry (\ref{xx}), one uses 
Eq.~(\ref{psieq}) to express $\psi$ through $\tau$ and Eq.~(\ref{veq}) to express $v$ through $\tau$.
As a result one arrives at the following action
\be
L={M_{Pl}^2\over 2}\l -{4\over m_0^2}\l\d_i^2\tau\r^2-{8\over m_1^2}(\d_0\d_i\tau)^2+
4\d_0\d_i\tau\d_0\d_i\sigma+6(\d_0\tau)^2+2(\d_i\tau)^2\r\;.
\ee
Variation of this action with respect to $\sigma$ implies that $\tau=0$
 if both $\omega$ and $p$ are non-zero.
Then variation with respect to
 $\tau$ gives $\sigma=0$. So mixing with gravity does not introduce new 
propagating modes in
this case as well.
\section{Discussion}
\label{final}
In this final section let us first briefly discuss what kind of cosmology can one expect 
in models of massive gravity discussed above. A hope that long scale
modification of gravity may
explain the observed
acceleration of  the cosmological expansion rate, or, 
at least, relate it to  effects testable
at shorter scales (such as the anomalous precession of the 
Moon~\cite{Lue:2002sw,Dvali:2002vf}) was the main driving force for the interest to 
massive gravity and its brane world analogues during the last few years. 
The common lore, supported
by the brane world models (e.g., by the DGP model~\cite{Dvali:2000hr}) is that modification
of the gravitational potential at the distance scale $r_c$ is related to the modification
of the Friedmann equation at the value of the Hubble expansion rate $H$ of order $r_c^{-1}$.

An example of the ghost condensate suggests that in four-dimensional models of 
massive gravity situation may be completely different. Indeed, at the linear
level this model is characterized by distance scale $r_c$ and time scale $t_c$,
such that gravitational potential of static sources is strongly modified at
distance and time scales larger than $r_c$ and $t_c$ correspondingly.
On the other hand, the homogeneous Friedmann equation in this model is completely equivalent
to the standard one (modulo, possibly, the presence of additonal dust-like component, 
with relative density depending on the initial conditions for the scalar field) for 
arbutrary value of the Hubble rate.

To illustrate that this is not a peculiar property of the 
ghost condensate, let us consider Lorentz-violating gravity with function $F$ (see
Eq.~(\ref{partinv})) of the following form 
\be
\label{cosmaction}
F=\Lambda^4F(XY^3)\;.
\ee
In the unitary gauge this means, that we are adding to the Einstein action 
the following non-covariant term
\[
\int d^4x\Lambda^4\sqrt{-g}F(g^{00}(\Tr g^{ij})^3)\;.
\]
We will comment later what was the reason to assume this particular form of
the action.  At the moment, let us just note that independence of function $F$
on $Z$ can be ensured, e.g., by time inversion symmetry. Also, for the
solutions we discuss here $V^i=0$, so possible dependence of $F$ on $V^i$ would not
affect our discussion.  Finally, there is a global symmetry
\[
\phi^0\to \lambda\phi^0\;,\;\;\phi^i\to\lambda^{-1/3}\phi^i
\]
protecting dependence of function $F$  on the combination
$XY^3$ at the quantum level.

Flat homogeneous cosmologies in massive gravity are described by the following ansatz for the 
metric and Goldstones,
\begin{gather}
\label{cosmansatz}
ds^2=dt^2-a^2(t)dx^idx^i\\
\phi^0=\phi^0(t)\\
\phi^i=\Lambda^2x^i
\end{gather}
In particular, $Y= 1/a^2$ and goes to zero as the Universe expands, so one may
suspect that effect of all masses but $m_0^2$ is negligible at late
times. We will see, however, that this is not necessarily the case.
Ansatz (\ref{cosmansatz}) automatically satisfies $\phi^i$ field equations, while 
$\phi^0$ equation reads as follows
\be
\label{phi0}
\d_t(a^3F'(XY^3)Y^3\d_t\phi^0)=\d_t\l F'(Xa^{-6})X^{1/2}a^{-3}\r=0\;.
\ee
A solution to this equation is 
\[
X=x_0 a^6
\]
or
\[
\phi_0(t)=x_0^{1/2}\int^tdta^3(t)\;.
\]
Combination $XY^3$ entering the Goldstone action remains constant for this solution.
On the first hand this implies that graviton masses are constant during the cosmological 
evolution. In particular, if some fine-tuning relation were true in the begining of expansion
it keeps to be true during the whole history of the Universe. For instance, it is 
straightforward to check, that action (\ref{cosmaction}) allows cosmological solutions in the
$m_0^2=0$ phase of massive gravity.

On the other hand, time-independence of $XY^3$ implies that contribution of the Goldstone
sector to the energy-momentum tensor is also constant. In other words, this contribution
is either zero or cosmological constant. So this is an example of the model where all
graviton modes are massive, but flat homogeneous solutions are the same as in the Einstein
gravity. Should one be disappointed by this fact? We believe that the answer is not.
This just implies that the value of the graviton mass may be quite large, affecting
growth of primordial perturbations, structure formation, or even gravitational dynamics
at the (super)galactic scales. 

Even more intriguingly, this opens up a possibility that the effects of
backreaction of small scale inhomgeneities on the expansion rate of the
Universe can be quite substantial in massive gravity. Actually, the
backreaction effects in conventional Einstein theory are not completely
understood, but it appeares that the likely conclusion is that they are very
small in the realistic Universe (see,
e.g. Refs.~\cite{Futamase:1996fk,Buchert:2001sa,Rasanen:2003fy,Kolb:2004am}
for a recent discussion).  However, this may be completely different in
massive gravity.  Indeed, in this case there is at least one extra length
scale in the problem --- inverse mass of the graviton. Now imagine the
Universe consisting of the heavy point-like galaxies at large distances from
each other. Then, when the distances between neighboring galaxies are larger
than the inverse graviton mass, the pairwise interactions between galaxies are
practically negligible (at least, in the $m_0^2=0$ phase, where the
gravitational potential is of the Yukawa type~\cite{Rubakov:2004eb}).  Is it
possible to use in this case the homogeneous approximation in cosmology? 
The answer to this question is not
clear to us.

An indirect
 hint that there may be a subtlety here is coming from the attempt to find closed or open
homogeneous cosmological solutions in the model (\ref{cosmaction}). In this case instead of 
a distance
between galaxies there is another extra length 
scale in the problem --- curvature radius\footnote{We thank Dominik Schwarz for suggesting us 
to look for open or closed solutions.}.
It is straightforward to see, however, that the 
corresponding ansatz does not satisfy all the equations (except the case of the de Sitter space,
when there are extra symmetries). This implies that closed or open Universe is necessarily
inhomogeneous on the scale of its curvature radius in massive gravity with action
(\ref{cosmaction}).

To conclude this brief discussion of the cosmology in massive gravity, it is
worth noting that time-independence of the graviton mass during the
cosmological expansion is by no means a general property of massive gravity.
Actually, the particular form of the action (\ref{cosmaction}) was derived from
the requirement that graviton mass is constant. The study of cosmological
consequences of the time-dependent graviton mass may provide new surprises.

Finally, let us stress that there is a large number of problems not addressed
in the current paper.  First of all, there are still a lot of questions about
stability of massive gravity.  For instance, our analysis of mixing with gravity in
section~\ref{vDVZ} covers only the phases $m_0^2=0$ (where it only slightly
extends that performed in Ref.~\cite{Rubakov:2004eb}), $m_1^2=0$ and a point
with enhanced symmetry in the phase with $m_2^2=m_3^2$. More detailed analysis
of the phases with $m_2^2=m_3^2$ and $m_2^2=0$ with negative $m_1^2$ is needed
to decide which regions of parameters identified in section~\ref{explicit}
actually correspond to tractable low-energy effective theories with cutoff
$\Lambda$. Also, it is necessary to include higher derivative terms in this
analysis. A particulary interesting question is whether two new UV insensitive
theories found here may avoid Jeans-like instability present in the ghost
condensate case. Furthermore, the analysis of stability of non-linear solutions
(on the first hand, cosmological ones and massive analogues of the Schwartzchild metric)
is needed to conclusively decide on the phenomenological acceptability of
models discussed here.

Also, in the current paper we concentrated on the issues of consistency and stability
 and leave aside
the analysis of phenomenology of modification of gravity in models studied here. 
Definitely, one of the most pressing questions is what are the actual limits
on the graviton mass parameters. 
It seems likely, that to obtain an accurate answer to this
question a detailed study of growth of primordial perturbations and structure
formation in massive gravity is needed.

Another circle of questions is what are the limits on the direct couplings
between Goldstone sector and Standard Model fields and what are the
implications of massive gravity for inflation (see,
Refs.~\cite{Arkani-Hamed:2004ar,Arkani-Hamed:2003uz} where these issues were
studied in the case of ghost condensate). We hope to address these and other
related questions in future.

\section*{Acknowlegements}
We appreciate stimulating discussions and correspondence with Peter Tinyakov,
Igor Tkachev, Misha Shaposhnikov, Dominik Schwarz, and Valery Rubakov. Special
thanks are to Riccardo Rattazzi for numerous fruitful discussions and
collaboration on some of the subjects discussed here. We thank organizers of the 9th Summer Institute
at the Gran Sasso Laboratory (in particular, Zurab Berezhiani), where part of this work was done, for a 
kind hospitality.

\end{document}